\documentclass[%
superscriptaddress,
preprint,
amsmath,amssymb,
aps,
]{revtex4-1}

\usepackage{color}
\usepackage{graphicx}
\usepackage{dcolumn}
\usepackage{bm}
\usepackage{enumitem}
\usepackage{ulem}

\begin {document}

\title{Detection of Transition Times from Single-particle-tracking Trajectories}

\author{Takuma Akimoto}
\email{takuma@rs.tus.ac.jp}
\affiliation{%
Department of Physics, Tokyo University of Science, Noda, Chiba 278-8510, Japan
}%

\author{Eiji Yamamoto}
\affiliation{%
Graduate School of Science and Technology, Keio University, Yokohama, Kanagawa 223-8522, Japan
}%


\date{\today}

\begin{abstract}
In heterogeneous environments, the diffusivity is not constant but changes with time.
It is important to detect changes in the diffusivity from single-particle-tracking trajectories in experiments.
Here, we devise a novel method for detecting the transition times of the diffusivity from trajectory data.
A key idea of this method is the introduction of a characteristic time scale of the diffusive states, which is obtained by 
a fluctuation analysis of the time-averaged mean square displacements.
We test our method in silico by using the Langevin equation with a fluctuating diffusivity. 
We show that our method can successfully detect the transition times of diffusive states and obtain the diffusion coefficient as a function of time. 
This method will provide a quantitative description of the fluctuating diffusivity in heterogeneous environments and can be applied to
 time series with transitions of states.
\end{abstract}

\maketitle



The mean square displacement (MSD) is one of the most popular observables for quantifying the diffusivity. In Brownian motion, 
the MSD increases linearly with time, and the diffusivity can be quantified by the slope of the MSD, i.e., the diffusion coefficient. 
The diffusion coefficient is determined by the surrounding environment including the viscosity of the medium
and the properties of the diffusing particle, e.g., the shape of the Brownian particle.
When there are no fluctuations in the properties of the surrounding environment and the diffusing particle, 
no intrinsic differences arise between the diffusivities for short-time and long-time measurements 
except for fluctuations of the diffusivity due to the finite measurement times.

In heterogeneous environments such as amorphous materials and living cells, diffusion often becomes anomalous; 
that is, the MSD does not increase linearly with time~\cite{Scher1975, Golding2006, ManzoTorreno-PinaMassignanLapeyreLewensteinGarcia2015}. 
The local diffusivities in 
these environments are highly heterogeneous. These heterogeneities are sometimes static or fluctuating. 
For example, the charge transport in amorphous materials \cite{Scher1975} as well as the diffusion of proteins on DNA \cite{Graneli2006,Wang2006}
can be modeled by a quenched trap model, where a random 
walker jumps in static random energy landscape~\cite{bouchaud90}.
In other words, the characteristic time scale of a change in the energy landscape is much longer than that of random walkers.
On the other hand, in supercooled liquids, mobile and immoblie particles are distributed in space, and the diffusive properties (mobile and immobile properties) will change with time, i.e., dynamic heterogeneity~\cite{Yamamoto-Onuki-1998, Yamamoto-Onuki-1998a,Richert-2002}. 
Moreover, transmembrane proteins~\cite{SergeBertauxRigneaultMarguet2008, Weron2017} and 
membrane-bound proteins on biological membranes~\cite{YamamotoAkimotoKalliYasuokaSansom2017} 
exhibit a temporally heterogeneous diffusivity. In these systems, the diffusivity for short-time measurements is intrinsically different from that for long-time measurements.

One of the most important issues in heterogeneous environments is to uncover the local diffusivity from single-particle trajectories. 
However, there is a crucial difficulty in extracting the local diffusivity in both spatially and temporally heterogeneous environments. 
In particular, one cannot know the boundaries of regions with the same diffusivities and transition times when the diffusive states change in spatially and temporally heterogeneous environments, respectively.
In previous studies, maximum likelihood estimators were proposed to determine the dynamic changes of the diffusivity~\cite{MontielCangYang2006, KooMochrie2016}, where the key idea is to detect the transition times when the diffusivity changes drastically. 
However, an empirical  parameter is necessary to implement the method.

The detection of the transition times is also important in state-transition processes, e.g., channel gating~\cite{WangVafabakhshBorschelHaNichols2016}, 
the conformational transition of proteins~\cite{ChungMcHaleLouisEaton2012},  the rotation of F1-ATPase~\cite{NojiYasudaYoshidaKinositaJr1997}, 
the fluorescence of quantum dots~\cite{StefaniHoogenboomBarkai2009}, and nanopore sensing of single molecules~\cite{HoworkaSiwy2009}.
A method for detecting transition times using only trajectories without prior knowledge and empirical parameters is desired in time-series analysis.

Here, we devise an estimation method for characterizing the short-time diffusivity from trajectory data without knowing the transition times of 
the diffusive states.
In our method, there are no parameters that are determined empirically.
Thus, our method can be applied when many single-particle-tracking trajectories are obtained.
We show that our method can successfully detect the transition times of the diffusivity and estimate the local diffusivity in 
the (overdamped) Langevin equation with  a fluctuating diffusion coefficient.


We assume that there are many trajectories for the same system and that the system can be described by the 
Langevin equation with a fluctuating diffusivity  (LEFD):
\begin{equation}
 \frac{d\bm{r}(t)}{dt} = \sqrt{2 D(t)} \bm{w}(t),
 \label{LEFD}
\end{equation}
where $\bm{r}(t)$ is the position of a particle at time $t$, $\bm{w}(t)$ is $d$-dimensional white Gaussian noise with $\langle \bm{w}(t)\rangle=0$ and $\langle w_i(t)w_j(t')\rangle = \delta_{ij}\delta (t-t')$, and $D(t)$ is the diffusion coefficient at time $t$, which is a stochastic process independent of $\bm{w}(t)$.
Although we do not assume any condition on $D(t)$, i.e., the diffusion coefficient may be non-Markov and depend on the position $\bm{r}(t)$, we assume
 that the variance of $D(t)$ is sufficiently large. In particular, it is much greater than the variance of the diffusion coefficients obtained by 
 the time-averaged MSD defined by Eq.~(\ref{tamsd}) when the measurement time $t$ 
 is the same as the characteristic time scale of the diffusive state.

In our setting, we do not know 
\begin{enumerate}[label=\roman*)]
\item the number of diffusive states and
\item the time scales of the diffusive states.
\end{enumerate}
This is because we do not know the transition times when a diffusive state changes in single-particle-tracking trajectories.
This is one of the most difficult issues when estimating the fluctuating diffusivity. 
To overcome this difficulty, we apply a fluctuation analysis of the time-averaged MSDs to obtain a characteristic time scale of the diffusive states. 
The time-averaged MSD is defined as
\begin{equation}
\overline{\delta^{2}(\Delta;t)} = \frac{1}{t-\Delta} \int_0^{t-\Delta} \{\bm{r}(t'+\Delta) - \bm{r}(t')\}^2dt' .
\label{tamsd}
\end{equation}
To characterize the fluctuations in the time-averaged MSDs, we use the relative standard deviation (RSD) of the time-averaged MSDs, defined as 
\begin{equation}
\Sigma(t;\Delta) \equiv \frac{\sqrt{\langle [\overline{\delta^{2}(\Delta;t)} -
 \langle \overline{\delta^{2}(\Delta;t)} \rangle]^{2} \rangle}}
 {\langle \overline{\delta^{2}(\Delta;t)} \rangle},
\end{equation}
as a function of the measurement time $t$ ($\Delta$ is fixed).
This type of quantity is widely used to investigate the ergodic property \cite{He2008, Deng2009} as well as the characteristic time of the system \cite{Akimoto2011, Miyaguchi2011a, Uneyama2012, Uneyama2015, Miyaguchi2016}.
In fact, the RSD analysis provides the longest relaxation time in 
the reptation model, which is a model of entangled polymers \cite{Uneyama2012, Uneyama2015}.

When $D(t)$ is a stationary stochastic process, i.e., the characteristic time of the stochastic process $D(t)$ is finite, the general formula for the RSD is derived as~\cite{Uneyama2015}
\begin{equation}
 \label{rsd_square_final}
  \Sigma^{2}(t;\Delta)
  \approx \frac{2}{t^{2}} \int_{0}^{t} ds \,
  (t - s) \psi_{1}(s) ,
\end{equation}
where $\psi_1(t)$ is the normalized correlation function of the diffusion coefficient, i.e., $\psi_1(t) \equiv (\langle D(t) D(0) \rangle - \langle D \rangle^2)/\langle D
\rangle^2$.
Thus, if the relaxation time of the system is $\tau$ (roughly speaking, the correlation function decays as $\psi_1(t) \propto e^{-t/\tau}$), the asymptotic form of the RSD becomes 
\begin{equation}
 \label{rsd_square_asymptotic}
  \Sigma^{2}(t;\Delta)
  \approx
  \begin{cases}
   \displaystyle
   \psi_{1}(0) & (t \ll \tau) , \\
      \displaystyle
   \frac{2}{t}  \int_{0}^{\infty} dv \, \psi_{1}(v) & (t \gg \tau) .
  \end{cases}
\end{equation} 
Therefore, $\tau$ is obtained by the crossover time from the plateau to the $t^{-1/2}$ decay in the RSD. In particular, 
when the correlation function decays exponentially, the crossover time $\tau_c$ in the RSD is given by $\tau_c \cong 2\tau$. 
From many single-particle-tracking trajectories, one can calculate the time-averaged MSDs.
Taking the ensemble average of the time-averaged MSDs gives us the RSD.
In this way, one can obtain the characteristic time scale of $D(t)$ from single-particle-tracking trajectories.
  
Here, we devise a novel method to detect the changes in states from a single-particle-tracking trajectory.
First, we define the time-averaged diffusion coefficient (TDC) at time $t$ by
\begin{equation}
D(t;\Delta,T) \equiv \frac{ \int_t^{t+T-\Delta}\{\bm{r}(t'+\Delta) - \bm{r}(t')\}^2dt'}{2d\Delta(T-\Delta)}.
\end{equation}
There are two parameters, $\Delta$ and $T$, in the TDC.
We set $\Delta$ as the minimal time step of the trajectory; thus, it is not necessary to tune this parameter. 
On the other hand, we have to tune the parameter  $T$ by introducing a tuning parameter $a$ as $T = a \tau_c$.
Since $\tau_c$ is of the same order as the system's characteristic time, $a$ can be smaller than one. 
In what follows, we use $a=0.1$.

Second, using the effective diffusion coefficient $D_{\rm eff}$, which is obtained by the ensemble average of the time-averaged MSD, i.e., $\langle \overline{\delta^{2}(\Delta;t)} \rangle = 2dD_{\rm eff}\Delta$, we define the crossing points $c_i$ as the points at which
the TDC crosses $D_{\rm eff}$, 
i.e., $D(c_i; \Delta,T) < D_{\rm eff}$ and $D(c_i +\Delta t; \Delta,T) > D_{\rm eff}$ or $D(c_i; \Delta,T) > D_{\rm eff}$ and 
$D(c_i +\Delta t; \Delta,T) < D_{\rm eff}$, 
 satisfying $c_{i+1} - c_i > T $, where $\Delta t$ is the time step of the trajectories (see Fig.~\ref{LEFD_two_states}A).
Note that the crossing points are not exact points representing changes in the diffusive states because 
different diffusive states coexist in a time window $[t, t+T-\Delta]$ of $D(t;\Delta,T)$. 
Therefore, we define the transition time as $t_i \equiv c_i + T/2$. 
The term $T/2$ is not exact when the threshold is not at the middle of two successive diffusive states.
If only one transition occurs in the time interval  $[t_i, t_{i+1}-\Delta]$, which is a physically reasonable assumption, the transition times represent 
the points of changes in the diffusive states.
Note that some transition times of the diffusive states will be still missing.

To correct the transition times obtained above, we test whether successive diffusive states are significantly different.
Since we know the transition times of the diffusive states, we can estimate the diffusion coefficient in the time interval $[t_i, t_{i+1}]$: 
the diffusion coefficient of the $i$th diffusive state is given by
 \begin{equation}
\overline{D}_i \equiv \frac{ \int_{t_i}^{t_{i+1}-\Delta}\{\bm{r}(t'+\Delta) - \bm{r}(t')\}^2dt'}{2d\Delta(t_{i+1}-t_i-\Delta)}.
\label{DC_i}
\end{equation}
Since we consider a situation that $T$ is sufficiently large ($T/\Delta t >30$), 
fluctuations of $\overline{D}_i$ can be approximated as a Gaussian distribution by the central limit theorem. 
According to a statistical test, the $i$th and $j$th states can be considered as the same state if there exists $D$ such that both the $k=i$ and $k=j$ states satisfy
\begin{equation}
 D - \sigma_k Z \leq \overline{D}_k \leq D + \sigma_k Z,
\label{statistical_test}
\end{equation}
where  $\sigma_{k}^2$ is the variance of the TDC with the time window $t_{k+1}-t_k$ and the diffusion coefficient $D$, which is given by $\sigma_{k}^2\equiv \frac{4D^2 \Delta }{3(t_{k+1}-t_k)}$, and $Z$ is determined by the level of statistical significance, e.g., $Z=1.96$ when the $p$-value is 0.05.
Therefore, the transition times can be corrected if the two successive diffusion states are the same.
We repeat this procedure: Eq.~(\ref{DC_i}) will be calculated again after correcting the transition times $t_i$, and the above test
will be repeated to correct the transition times.

Furthermore, one can improve the transition times by changing the thresholds around the transition times. The detailed procedure and  
 flowchart of our method are given in the Supplemental Material \cite{SM}.

\begin{figure}
\includegraphics[width=75 mm, bb= 0 0 219 272]{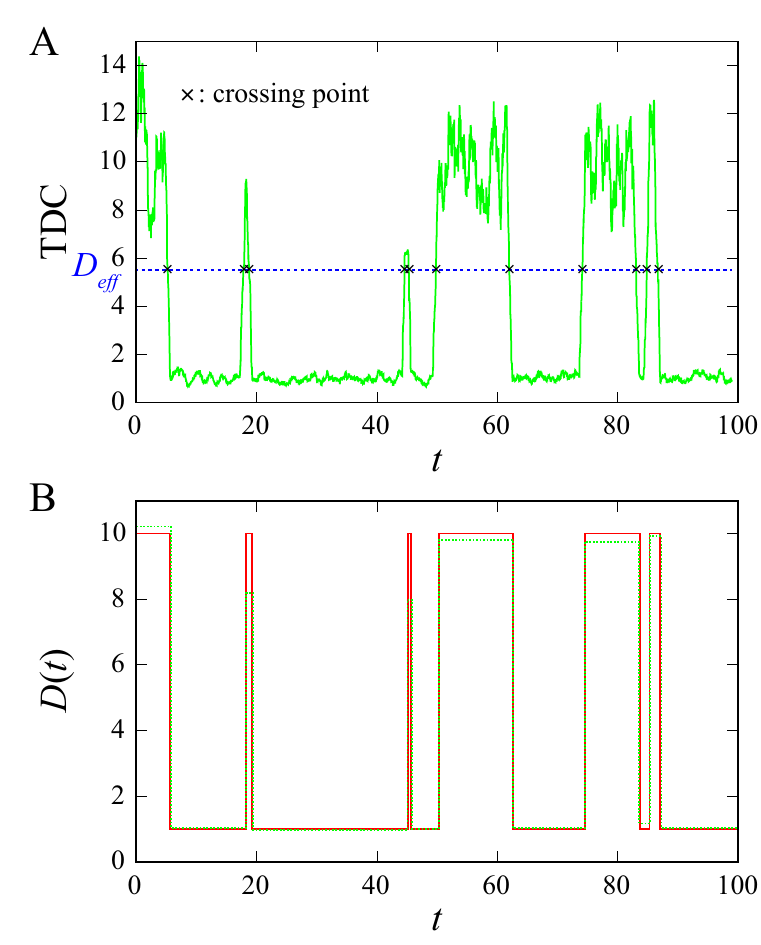}
\caption{Estimation of the diffusivities from a trajectory of a two-state LEFD model.
The sojourn-time distributions for the two states follow the exponential distribution with the same relaxation time ($\tau=10$).
(A)~Time-averaged diffusion coefficient $D(t;\Delta,T)$ as a function of time $t$ ($\Delta=0.01$ and $T=1$).
A trajectory is generated by the LEFD, and the diffusion coefficient $D(t)$ takes two values, $D=1$ and $10$.
The effective diffusion coefficient is given by $D_{\rm eff}=5.5$.
(B)~Diffusion coefficient $\overline{D}_i$ of the $i$th diffusive state with the true diffusion  coefficient $D(t)$ as a function of time $t$.
The green dashed and red solid lines represent the obtained and true diffusion coefficients, respectively.}
\label{LEFDtwo}
\end{figure}

Here, we test our method with the trajectories of three different LEFD models, where the number of diffusive states is two, three, and uncountable. 
The crossover times in the RSD are finite for all models.

In the Langevin equation with the two-state diffusivity, 
Fig.~\ref{LEFDtwo}B shows the diffusion coefficient obtained by our method.
Almost all diffusive states can be classified into two states according to the condition (\ref{statistical_test}) with $Z=1.96$. 
Moreover, the deviations in the transition times from the actual transition times are within 0.25.
Thus, we successfully extract the underlying diffusion process $D(t)$ from a single trajectory after obtaining the characteristic time scale of the diffusive states. 

\begin{figure}
\includegraphics[width=75 mm,bb= 0 0 209 132]{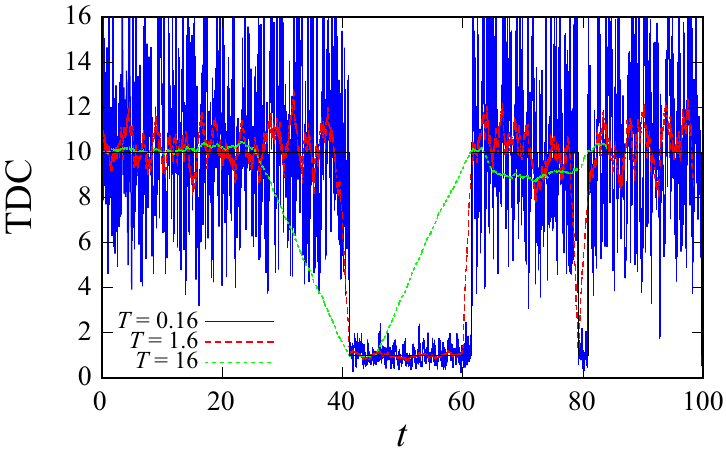}
\caption{Time-averaged diffusion coefficient $D(t;\Delta,T)$ as a function of time $t$ for different time windows $T$ ($\Delta=0.01$).
 The sojourn-time distributions for both states are exponential distributions.
 A trajectory is generated by the LEFD, and the diffusion coefficient $D(t)$ takes two values, 
$D_1=10$ and $D_2 =1$, with relaxation times of 40 and 10, respectively, where  $\tau_c =16$ and $D_{\rm eff}=8.2$.}
\label{tdc_two_state}
\end{figure}

We introduce different relaxation times in the two sojourn-time distributions (we use the exponential distribution for both sojourn-time distributions) and
examine the effects of the tuning parameter.
Figure~\ref{tdc_two_state} shows the TDCs for different tuning parameters $a=1$, $0.1$, and 0.01, corresponding 
to $T=16$, 1.6, and 0.16, respectively. 
As clearly seen, when the tuning parameter is small, the fluctuations in the TDC become large.
Therefore, inaccurate transition times may be detected when $a$ is too small.
On the other hand, the actual transition times may not be detected when $a$ is too large.
In fact, the transition times around $t=80$ cannot be detected in the case of the green dotted line ($a=1$). 
As a result, the tuning parameter can be set to $a=0.1$ or between 0.1 and 0.01.

\begin{figure}
\includegraphics[width=75 mm,bb= 0 0 219 278]{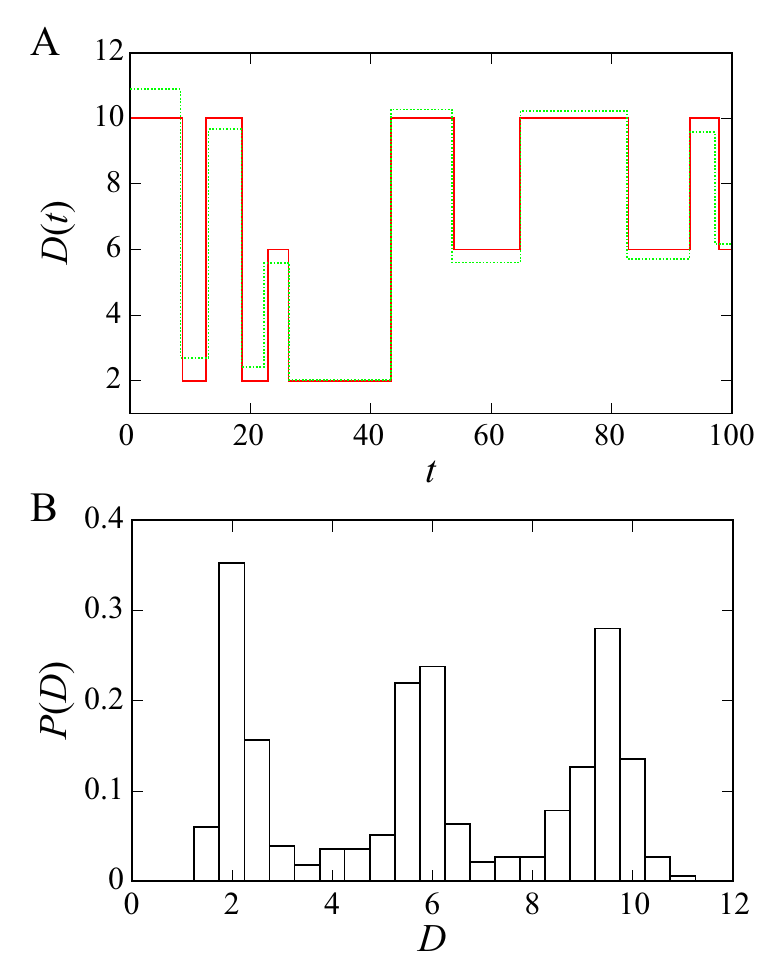}
\caption{(A) Diffusion coefficient $D_i$ of the $i$th diffusive state with the true diffusion 
coefficient $D(t)$ as a function of time $t$. A trajectory is generated by the LEFD, and  $D(t)$ takes three values, 
$D_1=10, D_2=6$, and $D_3 =2$, with a relaxation time of 10, where the transition probability is given by $p_{ij}=1/2$ ($i\ne j$),  
and $D_{\rm eff}=6$.
The green dashed and red solid lines represent the obtained and true diffusion coefficients, respectively. 
(B) Histogram of the estimated diffusion coefficient. The parameters are the same as those in (A).}
\label{tdc_LEFD3}
\end{figure}

Next, we analyze the LEFD with the three-state diffusivity. The sojourn-time distributions are exponential distributions, and their 
relaxation times are the same in each state ($\tau=10$).
For the three-state LEFD, one can obtain several diffusive states from a single trajectory by our method after calculating the crossover time in 
the RSD using many trajectories.
Figure~\ref{tdc_LEFD3} shows the diffusion coefficient obtained by our method, where we revised the threshold using 
the procedures described in the Supplemental Material \cite{SM}. As shown in Fig.~3A, the transition times are correctly detected. 
Moreover, almost all diffusive states belong to the three diffusive states using $Z=1.96$, and the distribution of the estimated 
diffusion coefficients has three peaks corresponding to the exact diffusion coefficient (see Fig.~3B). 

Finally, we apply our method to a diffusion process with an uncountable number of diffusive states. In particular, we use  the annealed transit time model (ATTM)~\cite{MassignanManzoTorreno-PinaGarcia-ParajoLewensteinLapeyre2014,AkimotoYamamoto2016a}.
The ATTM was proposed to describe heterogeneous diffusion in living cells~\cite{MassignanManzoTorreno-PinaGarcia-ParajoLewensteinLapeyre2014,ManzoTorreno-PinaMassignanLapeyreLewensteinGarcia2015}.
The diffusion process is described by the LEFD where $D(t)$ is coupled to the sojourn time. 
When  the sojourn time is $\tau$, the diffusion coefficient is given by $D_{\tau} = \tau^{\sigma - 1} (0 < \sigma < 1)$. Here, we assume that
the sojourn-time distribution follows an exponential distribution $\rho(\tau) \sim \exp (- \tau / \langle \tau \rangle ) / \langle \tau \rangle$.
One can obtain  $\tau_c$  by the RSD analysis~\cite{AkimotoYamamoto2016a}.

\begin{figure}
\includegraphics[width=75 mm,bb= 0 0 219 286]{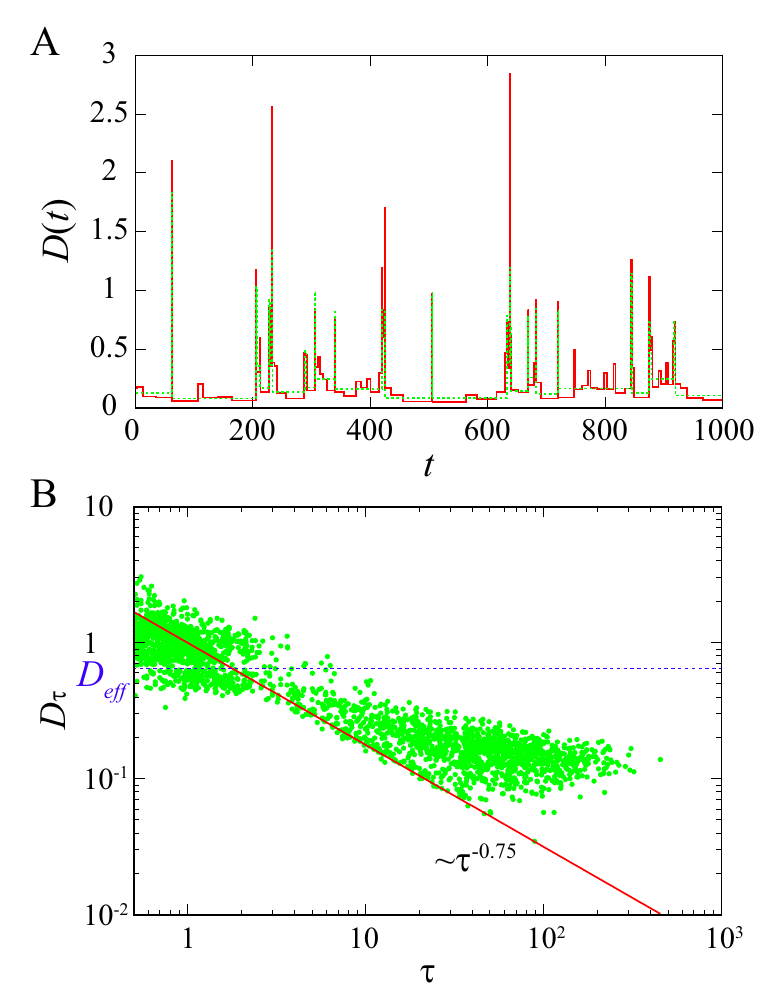}
\caption{Estimation of the diffusivity of the ATTM.
(A)~Diffusion coefficient $D_i$ of the $i$th diffusive state with the true diffusion coefficient $D(t)$ 
as a function of time $t$.
A trajectory is generated with $\langle \tau \rangle = 10$ and $\sigma = 0.25$.
The green dashed and red solid lines represent the obtained  and true diffusion coefficients, respectively.
(B)~Relation between the estimated diffusion coefficients and the sojourn times.}
\label{ATTM}
\end{figure}

Figure~\ref{ATTM}A shows the diffusion coefficient obtained by our method.
Because the variance of $D(t)$ is not large, the transition times are not correctly detected compared with the other two models.
However, the transition times for the highly diffusive states can be detected correctly. 
Moreover, Fig.~\ref{ATTM}B shows the relation between the obtained diffusion coefficient and the sojourn times, 
which exhibits a power-law relation  $D_\tau  =\tau^{\sigma - 1}$.
Therefore, our method can also be applied to systems with an uncountable number of diffusive states.

Diffusivity changes with time in temporally/spatially heterogeneous environments such as cells and supercooled liquids.
It is difficult to estimate such a fluctuating diffusivity from single-particle trajectories because one does not have information about the
transition times when the diffusivity changes.
In this paper, we have proposed a new method for detecting the transition times from single trajectories.
Our method is based on a fluctuation analysis of the time-averaged MSD to extract information on the characteristic time scale 
of the system. 
We have applied this  method to three different diffusion processes, i.e., the LEFD with two states, the LEFD with three states, and the ATTM, which 
has an uncountable number of diffusive states.
Our method successfully extracts the transition times of the diffusivities and estimates the fluctuating diffusion coefficients in the three models.
Since our method can be conducted with single-particle trajectories, the application will be useful and of importance in experiments.
Furthermore, a slight modification of this method will be also applied to the time series of state-transition processes.

E.Y. was supported by an MEXT (Ministry of Education, Culture, Sports, Science and Technology) Grant-in-Aid for the ``Building of Consortia for the Development of Human Resources in Science and Technology.''



%

\end{document}